\documentclass[twocolumn,  
showpacs,  
preprintnumbers,  
aps,  
prd,  
a4paper,  
superscriptaddress,  
nofootinbib,  
tightenlines,  
floats,floatfix  
]{revtex4}

\usepackage{amsmath}
\usepackage{amsfonts}
\usepackage{graphicx}

\def\be{\begin{equation}}
\def\ee{\end{equation}}
\def\ba{\begin{eqnarray}}
\def\ea{\end{eqnarray}}
\def\bs{\begin{subequations}}
\def\es{\end{subequations}}

\newcommand{\rd}{{\rm d}}

\def\R{\cal{R}}

\usepackage{color}

\newcommand{\nn}{\nonumber\\}
\newcommand{\p}[1]{(\ref{#1})}

\begin{document}

\preprint{KU-TP 009}
\preprint{hep-th/0610336}

\title{Realizing Scale-invariant Density Perturbations
in Low-energy Effective String Theory}

\bigskip

\author{Zong-Kuan Guo}
\email{guozk@phys.kindai.ac.jp}

\author{Nobuyoshi Ohta}
\email{ohtan@phys.kindai.ac.jp}
\affiliation{Department of Physics, Kinki University, Higashi-Osaka,
Osaka 577-8502, Japan}

\author{Shinji Tsujikawa}
\email{shinji@nat.gunma-ct.ac.jp}
\affiliation{Department of Physics, Gunma National College of
Technology, Gunma 371-8530, Japan}

\date{\today}

\begin{abstract}

We discuss the realization of inflation and resulting cosmological
perturbations in the low-energy effective string theory.
In order to obtain nearly scale-invariant spectra of density
perturbations and a suppressed tensor-to-scalar ratio,
it is generally necessary that
the dilaton field $\phi$ is effectively decoupled from gravity together
with the existence of a slowly varying dilaton potential.
We also study the effect of second-order corrections to the tree-level
action which are the sum of a Gauss-Bonnet term  coupled to $\phi$
and a kinetic term $(\nabla \phi)^4$. We find that it is possible to realize
observationally supported spectra of scalar and tensor perturbations
provided that the correction is dominated by the $(\nabla \phi)^4$
term even in the absence of the dilaton potential.
When the Gauss-Bonnet term is dominant,
tensor perturbations exhibit violent negative instabilities
on small-scales about a de Sitter background in spite of the fact
that scale-invariant scalar perturbations can be achieved.

\end{abstract}


\maketitle

\section{Introduction}

The constantly accumulating observational data including
WMAP \cite{WMAP,WMAP3}, SDSS \cite{SDSS} and
2dF \cite{2dF} have continued to confirm that primordial
density perturbations are adiabatic and nearly scale-invariant.
This is consistent with the prediction of inflationary
paradigm which is based upon a scalar field with
a slowly varying potential
(see Refs.~\cite{review,Lidsey97,BTW} for review).

So far there are many attempts to try to explain the origin of
density perturbations in the context of string theory.
For example, in the low-energy effective string theory, a dilaton field
$\phi$ coupled to a scalar curvature $R$ leads to a so-called
Pre-Big-Bang phase during which superinflation occurs
in the string frame \cite{PBB}.
In the Einstein frame, this corresponds to a contracting universe
driven by the kinetic energy of the dilaton.
In this case the spectrum of curvature perturbations is highly
blue-tilted ($n_{\cal R}=4$) \cite{Brustein94}, thus incompatible
with observations unless another scalar field such as axion
works as a curvaton \cite{curvaton}.

The ekpyrotic/cyclic scenarios \cite{ekp} also lead to a contracting
universe driven by a negative exponential potential.
The perturbation spectra generated in these scenarios have been
discussed by many authors (see e.g., \cite{ekpper}).
In order for a pre-bounce growing mode of a scale-invariant
Bardeen potential $\Phi$ to survive long after the bounce,
it is necessary that the pressure perturbation is directly proportional to
a Bardeen potential, but this mode is not present for any known
form of an ordinary matter \cite{Bozza}.
Hence it is still a challenging task to
realize scale-invariant density perturbations in
bouncing cosmologies.

Recently a number of authors have discussed the spectra of
density perturbations in string-gas cosmology \cite{stringgas}.
The authors in Refs.~\cite{stringgas,Biswas} argued that a nearly
scale-invariant perturbation may be obtained in the Hagedorn
regime seeded by a gas of strings in thermal equilibrium.
In recent papers it was realized that the original models
in Refs.~\cite{stringgas} lead to a highly blue-tilted spectrum
($n_{\cal R}=5$) unless the dilaton is fixed in the
Hagedorn phase \cite{Robert,Kaloper}.
The coupling of the dilaton to gravity nontrivially changes
the cosmological evolution and resulting density perturbations.

The above examples show that it is generally difficult to
obtain nearly scale-invariant cosmological perturbations
without using slow-roll inflation driven by a potential
energy of a scalar field minimally coupled to gravity.
In the low-energy effective string theory,
the potential of the dilaton field is absent at the tree level, thus
posing a difficulty to get slow-roll inflation.
Moreover when the dilaton is coupled to a Ricci scalar,
this has a significant effect on the spectrum of density perturbations
as it happens in Pre-Big-Bang and string-gas cosmologies.
It is known that in string theories there are higher-order
quantum corrections, which can alter these results.
However, the inflationary solutions and the resulting cosmological perturbations
have not been studied much when these higher-order effects are present.
It is thus important to examine what is the physical
significance of these quantum corrections.

In this paper we study the possibility to achieve observationally
supported cosmological perturbations in the low-energy string effective
action with a dilaton coupling $F(\phi)R$.
Since both curvature and tensor metric perturbations are invariant
under a conformal transformation in the absence of higher-order
corrections, the demand for realizing
scale-invariant spectra in the string frame and a suppressed
tensor-to-scalar ratio corresponds to obtaining
slow-roll inflation in the Einstein frame.
This then requires the presence of a dilaton potential
together with the condition that the dilaton is effectively
decoupled from gravity when the perturbations on
cosmologically relevant scales are generated.

We also take into account the higher-order corrections
which are the sum of the contributions of a
Gauss-Bonnet (GB) term and a kinetic term $(\nabla \phi)^4$.
When the $(\nabla \phi)^4$ term is dominant,
we find that it is in fact possible to realize
observationally supported density perturbations
even if the dilaton potential is absent.
The reason why this ``kinetic inflation'' \cite{kinf} can be successful is
that inflation of a slow-roll type is realized in the Einstein frame,
unlike the Pre-Big-Bang and ekpyrotic/cyclic models.

Inflation can be also realized when the GB term is present.
However this does not necessarily mean that the models
are compatible with observations.
In fact, with several different choices of the coupling $F(\phi)$,
we show that either of the following two cases occurs
when the GB term is dominant:

(i) The power spectra of curvature and tensor perturbations
are far from scale-invariant ones, or

(ii) While the spectra of scalar perturbations can be nearly
scale-invariant, tensor perturbations exhibit negative
instabilities on small scales which invalidates the assumption
of linear perturbations.

Thus the GB-dominated inflation is generally problematic to generate
observationally supported density perturbations from quantum
fluctuations.
However, we find that if the GB term is subdominant compared with the
higher-order kinetic term, it is possible to obtain the density perturbations
compatible with observations.

This paper is organized as follows.
In Sec.~\ref{perturbation} the formulae
for the spectra of scalar and tensor metric perturbations are
presented for the low-energy string effective action
with higher-order correction terms.
In Sec.~\ref{einstein} we discuss the transformation of the action to
the Einstein frame and show the equivalence of perturbation
spectra in both the string and Einstein frames.
In Sec.~\ref{dilaton} we study the possibility to obtain
scale-invariant cosmological perturbations in dilaton gravity
without higher-order corrections.
We consider general models without restricting the Pre-Big-Bang
cosmology in the latter half of Sec.~\ref{dilaton}.
We then investigate cosmological
perturbations in the theories with higher-order kinetic
terms $(\nabla \phi)^4$ in Sec.~\ref{kinetic}.
Models with higher-order corrections which are
the sum of the GB term and the $(\nabla \phi)^4$ term are discussed
in Sec.~\ref{correctionsec}.
Our new findings are 
the results presented in Sec.~\ref{kinetic} and Sec.~\ref{correctionsec}
together with the latter half of Sec.~\ref{dilaton}.
Sec.~\ref{conclusion} is devoted to conclusions.

\section{Cosmological perturbations for a general action}
\label{perturbation}

In this section, we first develop the formalism for analysing the density
perturbations for a system with higher-order corrections.
Specifically we consider the following general action
\begin{eqnarray}
\label{action}
S &= & \int {\rm d}^4 x\sqrt{-g} \Bigg[ \frac12 F(\phi)R
-\frac12 \omega (\phi) (\nabla \phi)^2 \nn
&& \hspace{20mm} -V(\phi)+{\cal L}_{c} \Bigg],
\end{eqnarray}
where $\phi$ is a scalar field with a potential $V(\phi)$,
$R$ is a Ricci scalar,
$F(\phi)$ and $\omega (\phi)$ are the functions of $\phi$.
In what follows, we use the unit $\kappa^2 \equiv
8\pi G=1$ with $G$ being gravitational constant.
The Lagrangian ${\cal L}_{c}$ represents the contribution
of higher-order corrections to be specified shortly.

In the weak coupling limit of the low-energy effective string theory,
where the string coupling $g_s^2 \equiv e^{\phi}$ is
much smaller than unity, the action
corresponds to $F(\phi)=e^{-\phi}$,
$\omega (\phi)=-e^{-\phi}$, and $V(\phi)=0$ \cite{PBB,Easther}.
We only consider a dilaton field $\phi$
by assuming that other modulus fields corresponding to
the size of extra dimensions are fixed through some
mechanism.\footnote{See Refs.~\cite{modulus} for the
the cosmological evolution in the presence of modulus fields.}
The higher-order corrections ${\cal L}_{c}$ are the
infinite sums of the series expansion with an expansion
parameter $\alpha'=l_s^2$, where $l_s$ is the string length.
We pick up lowest-order terms with which
the equations of motion are second-order
in fields \cite{higherpapers1,higherpapers}:
\begin{eqnarray}
\label{correction}
{\cal L}_c=-\frac12 \alpha' \xi (\phi)
\left[ c_{1} R_{\rm GB}^2 +c_2 (\nabla \phi)^4
\right]\,,
\end{eqnarray}
where $R_{\rm GB}^2=R^2-4R_{\mu\nu}R^{\mu\nu}+
R_{\alpha\beta\mu\nu}R^{\alpha\beta\mu\nu}$ is the GB term.
At the tree level, the coupling $\xi(\phi)$ and the coefficients
$c_1$ and $c_2$ are
\begin{eqnarray}
\xi(\phi)=\lambda e^{-\phi}\,, \quad
c_1=1\,, \quad c_2=-1\,,
\end{eqnarray}
where $\lambda=-1/4, -1/8, 0$ correspond to bosonic,
heterotic and Type II superstrings, respectively.
In what follows, we work in a unit $\alpha'=1$.

As the system enters a large coupling region characterized
by $g_s^2 \gtrsim 1$, it is expected that the forms of
the functions $F(\phi)$, $\omega (\phi)$ and $\xi(\phi)$
become more complicated than those given above.
Moreover the potential of the dilaton may
appear in order to stabilize the field.
Hence we work on the general action (\ref{action})
without restricting ourselves to the tree-level case.

In a flat Friedmann-Robertson-Walker (FRW) metric with
a scale factor $a$, we obtain the following
background equations \cite{CHC01,Hwang05,Gian}:
\begin{eqnarray}
& & H^2=\frac{1}{6F} \left( \omega \dot{\phi}^2
+2V-6H \dot{F}+2\rho_c \right),
\label{f1} \\
& & \dot{H}=\frac{1}{2F} \left( -\omega \dot{\phi}^2
+H \dot{F}-\ddot{F}-\rho_c-p_c \right),
\label{f2} \\
& & \ddot{\phi}+3H \dot{\phi}+\frac{1}{2\omega}
\left( \omega_{,\phi} \dot{\phi}^2-F_{,\phi}R+2V_{,\phi}
+T_{c} \right)=0\,,~~~
\label{f3}
\end{eqnarray}
where a dot represents the time derivative,
$\omega_{,\phi} \equiv \rd \omega /\rd \phi$, $H \equiv \dot{a}/a$,
and the correction terms are
\begin{eqnarray}
& & \rho_{c}=12 c_1 \dot{\xi}H^3-\frac32 c_2 \xi \dot{\phi}^4,\\
& & p_{c}=-4c_1 [ \ddot{\xi} H^2+2\dot{\xi}H (\dot{H}+H^2)]
-\frac12 c_2 \xi \dot{\phi}^4,\\
& & T_c=24 c_1 \xi_{,\phi} (\dot{H}+H^2)H^2 \nn
&& \hspace{10mm}-c_2 \dot{\phi}^2 (3\dot{\xi} \dot{\phi}+12 \xi \ddot{\phi}
+12 \xi \dot{\phi} H).
\end{eqnarray}

Let us consider the following perturbed metric about a FRW
background \cite{Bardeen}:
\begin{eqnarray}
\hspace*{-0.2em}\rd s^2 &=& - (1+2A)\rd t^2 +
2a\partial_iB \rd x^i\rd t
\nonumber\\
\hspace*{-0.2em}&& +a^2\left[ (1+2\psi)\delta_{ij}
+2\partial_{ij}E+2h_{ij}
\right] \rd x^i \rd x^j\,,
\end{eqnarray}
where $\partial_i$ represents the spatial partial derivative
$\partial/\partial x^i$ and
$\partial_{ij}=\nabla_i\nabla_j-(1/3)\delta_{ij}\nabla^2$.
We use lower case latin indices
running over the 3 spatial coordinates.
We note that $A$, $B$, $\psi$ and $E$ denote scalar metric
perturbations, whereas $h_{ij}$ represents tensor perturbations.
We define the so-called comoving
perturbation \cite{cocurvature}
\begin{eqnarray}
{\cal R} \equiv \psi-\frac{H}{\dot{\phi}}\delta \phi\,,
\end{eqnarray}
which is invariant under a gauge transformation.
The Fourier modes of curvature perturbations
satisfy \cite{CHC01,Hwang05}
\begin{eqnarray}
\label{Req}
\frac{1}{az^2} (az^2 \dot{\R})^\cdot
+c_{\R}^2 \frac{k^2}{a^2} {\R}
=0\,,
\end{eqnarray}
where $k$ is a comoving wavenumber and
\begin{eqnarray}
\label{z2def}
& & z^2= \frac{a^2 \left( \omega \dot{\phi}^2+
\frac{3(\dot{F}+Q_1)^2}{2F+Q_2}+Q_3\right)}
{\left(H+\frac{\dot{F}+Q_1}{2F+Q_2}\right)^2},
\\
\label{cRdef}
& & c_{\R}^2=1+\frac{Q_4+\frac{\dot{F}+Q_1}{2F+Q_2}Q_5+
\left(\frac{\dot{F}+Q_1}{2F+Q_2} \right)^2Q_6}
{\omega \dot{\phi}^2+
\frac{3(\dot{F}+Q_1)^2}{2F+Q_2}+Q_3},
\end{eqnarray}
with
\begin{eqnarray}
& &Q_1=-4c_1 \dot{\xi}H^2,\quad
Q_2=-8c_1 \dot{\xi}H,\quad
Q_3=-6c_2 \xi \dot{\phi}^4, \nonumber \\
& &Q_4=4c_2 \xi \dot{\phi}^4, \quad
Q_5=-16c_1 \dot{\xi} \dot{H}, \quad
Q_6=8c_1 (\ddot{\xi}-\dot{\xi}H).
\nonumber \\
\end{eqnarray}

Introducing a new variable, $v=z\R$, we find that
Eq.~(\ref{Req}) can be rewritten as
\begin{eqnarray}
\label{veq}
v''+\left( c_{\R}^2 k^2 -\frac{z''}{z}
\right) v=0\,,
\end{eqnarray}
where a prime represents a derivative with respect to
a conformal time $\tau=\int a^{-1} {\rm d}t$.
When the evolution of $z$ is given by $z \propto |\tau|^q$,
one has $z''/z=\gamma_{\R}/\tau^2$ with
$\gamma_{\R}=q(q-1)$.
In this case, if $c_{\R}^2$ is a positive constant, the solution
for Eq.~(\ref{veq}) can be written by using Hankel functions:
\begin{eqnarray}
v=\frac{\sqrt{\pi |\tau|}}{2}
\left[ c_1 (k) H_{\nu_{\R}}^{(1)} (c_{\R} k |\tau|)+
c_2 (k) H_{\nu_{\R}}^{(2)} (c_{\R} k |\tau|)
\right], \nonumber \\
\end{eqnarray}
where $\nu_{\R}=\sqrt{\gamma_{\R}+1/4}=|q-1/2|$.
We choose the coefficients to be $c_1=0$ and $c_2=1$,
so that positive frequency solutions in a Minkowski
vacuum are recovered in an asymptotic past.
Since $H_{\nu_{\R}}^{(2)} (c_{\R} k|\tau|) \to (i/\pi)
\Gamma(\nu_{\R}) (c_{\R} k|\tau|/2)^{-\nu_{\R}}$
for long wavelength perturbations ($c_{\R} k|\tau| \ll 1$),
the curvature perturbation after the Hubble radius
crossing is given by
\begin{eqnarray}
{\cal R}=i\frac{\sqrt{|\tau|}}{4z}
\frac{\Gamma (\nu_{\R})}{\Gamma (3/2)}
\left( \frac{c_{\R}k |\tau|}{2}\right)^{-\nu_{\R}}\,.
\end{eqnarray}

The spectrum of the curvature perturbation is defined
by ${\cal P}_{\cal R}=k^3 |{\cal R}|^2/2\pi^2$.
Then we find
\begin{eqnarray}
{\cal P}_{\cal R} &=& \frac{c_{\R}^{-2\nu_{\R}}}
{Q_{\R}}
\left(\frac{H}{2\pi}\right)^2
\left(\frac{1}{aH|\tau|}\right)^2
\left(\frac{\Gamma(\nu_{\R})}{\Gamma(3/2)}\right)^2
\left(\frac{k|\tau|}{2}\right)^{3-2\nu_{\R}}
\nonumber \\
&\equiv&
A_{\R}^2 \left(\frac{k|\tau|}{2}\right)
^{3-2\nu_{\R}},
\label{PS}
\end{eqnarray}
where
\begin{eqnarray}
Q_{\R} \equiv\frac{ \omega \dot{\phi}^2+
\frac{3(\dot{F}+Q_1)^2}{2F+Q_2}+Q_3}
{\left(H+\frac{\dot{F}+Q_1}{2F+Q_2}\right)^2}.
\end{eqnarray}
When $\nu_{\R}=0$, we have an additional ${\rm ln}\,(k|\tau|)$ factor.
{}From Eq.~(\ref{PS}) the spectral index of the power spectrum is
\begin{eqnarray}
n_{\cal R}-1=3-2\nu_{\R}=3-\sqrt{4\gamma_{\R}+1}\,.
\end{eqnarray}
Note that the scale-invariant spectrum ($n_{\cal R}=1$)
corresponds to $\gamma_{\R}=2$ (or $\nu_{\R}=3/2$).

We decompose tensor perturbations into eigenmodes of the
spatial Lagrangian, $\nabla^2 e_{ij}=-k^2e_{ij}$, with
scalar amplitude $h(t)$, i.e., $h_{ij}=h(t)e_{ij}$,
where $e_{ij}$ have two polarization states.
The Fourier modes of tensor perturbations
satisfy \cite{CHC01,Hwang05}
\begin{eqnarray}
\label{Teq}
\frac{1}{a^3Q_T} (a^3 Q_T \dot{h})^\cdot
+c_T^2 \frac{k^2}{a^2}h=0\,,
\end{eqnarray}
where
\begin{eqnarray}
Q_T=F+\frac{Q_2}{2} \,,\quad
c_T^2=1-\frac{4c_1 (\ddot{\xi}-\dot{\xi}H)}
{F-4c_1 \dot{\xi}H}\,.
\label{cT2}
\end{eqnarray}
Introducing new variables $z_T=a\sqrt{Q_T}$ and
$v_T=z_T h/2$, Eq.~(\ref{Teq}) can be rewritten as
\begin{eqnarray}
\label{vteq}
v_T''+\left( c_T^2 k^2 -\frac{z_T''}{z_T}
\right) v_T=0\,.
\end{eqnarray}
The power spectrum of tensor perturbations is defined by
${\cal P}_T=2\,k^3 |h|^2/2\pi^2$ because of  two polarization
states of the graviton.
If $c_T^2$ is a positive constant and the evolution of $z_{T}$
is given by $z_{T} \propto |\tau|^{q_T}$, we obtain
\begin{eqnarray}
{\cal P}_T &=& \frac{8c_T^{-2\nu_T}}{Q_T}
\left(\frac{H}{2\pi}\right)^2
\left(\frac{1}{aH|\tau|}\right)^2
\left(\frac{\Gamma(\nu_{T})}{\Gamma(3/2)}\right)^2
\left(\frac{k|\tau|}{2}\right)^{3-2\nu_T}
\nonumber \\
&\equiv&
A_T^2 \left(\frac{k|\tau|}{2}\right)
^{3-2\nu_T},
\label{PT}
\end{eqnarray}
where $\nu_{T}=\sqrt{\gamma_T+1/4}$ with
$\gamma_T=q_T (q_T-1)$.
The spectral index of the power spectrum is
\begin{eqnarray}
n_T=3-2\nu_T=3-\sqrt{4\gamma_T+1}\,.
\end{eqnarray}
The tensor-to-scalar ratio is given by
\begin{eqnarray}
\label{ratiodef}
r &\equiv& \frac{A_{T}^2}{A_{\R}^2}
\nonumber \\
&=&
8 \frac{\omega \dot{\phi}^2+\frac{3(\dot{F}+Q_1)^2}
{2F+Q_2}+Q_3}{\left(H+\frac{\dot{F}+Q_1}{2F+Q_2}
\right)^2 \left(F+\frac{Q_2}{2}\right)}
\frac{c_{\R}^{2\nu_{\R}}}{c_T^{2\nu_T}}
\left(\frac{\Gamma (\nu_T)}{\Gamma (\nu_{\R})}
\right)^2. \nonumber \\
\end{eqnarray}

We introduce the following quantities
\begin{eqnarray}
& &\epsilon_{1}=-\frac{\dot{H}}{H^2}\,, \quad
\epsilon_{2}=\frac{\ddot{\phi}}{H\dot{\phi}}\,, \quad
\epsilon_{3}=\frac{\dot{F}}{2HF}\,, \quad
\epsilon_{4}=\frac{\dot{E}}{2HE}\,,\nonumber \\
& &\epsilon_{5}=\frac{\dot{F}+Q_{1}}{H(2F+Q_2)}\,,\quad
\epsilon_{6}=\frac{\dot{Q}_T}{2H Q_{T}}\,,
\label{epdef}
\end{eqnarray}
where
\begin{eqnarray}
E \equiv \frac{F}{\dot{\phi}^2} \left[ \omega \dot{\phi}^2+
\frac{3(\dot{F}+Q_1)^2}{2F+Q_2}+Q_3 \right]\,.
\end{eqnarray}
Then the variable $z^2$ in Eq.~(\ref{z2def}) is given by
\begin{eqnarray}
z^2=\left(\frac{a\dot{\phi}}{H(1+\epsilon_5)}\right)^2
\frac{E}{F}\,.
\end{eqnarray}
In what follows, we specialize to the case in which
$\dot{\epsilon}_i$ and $\ddot{\epsilon}_i$ terms
vanish or the case in which they can be neglected compared to
other terms (like slow-roll inflation).
Since the conformal time is given by
$\tau=-1/(a H (1-\epsilon_1))$ in this case, we get
$z''/z=\gamma_{\cal R}/\tau^2$ with
\begin{eqnarray}
\gamma_{\cal R}=\frac{(1+\epsilon_1+\epsilon_2
-\epsilon_3+\epsilon_4)(2+\epsilon_2-\epsilon_3+\epsilon_4)}
{(1-\epsilon_1)^2}\,.
\end{eqnarray}
Then the spectral index of scalar perturbations is given by
\begin{eqnarray}
\label{nRex}
n_{\cal R}-1=3-\left| \frac{3+\epsilon_1+2\epsilon_2
-2\epsilon_3 +2\epsilon_4}
{1-\epsilon_1} \right|\,.
\end{eqnarray}
The variable $z_{T}$ for tensor perturbations
satisfies the relation $z_T''/z_T=\gamma_T/\tau^2$ with
\begin{eqnarray}
\gamma_{T}=\frac{(2-\epsilon_1+\epsilon_6)(1+\epsilon_6)}
{(1-\epsilon_1)^2}\,.
\end{eqnarray}
Hence the spectral index of tensor perturbations is
\begin{eqnarray}
\label{nTex}
n_{T}=3-\left| \frac{3-\epsilon_1+2\epsilon_6}
{1-\epsilon_1} \right|\,.
\end{eqnarray}

When the conditions $|\epsilon_i| \ll 1$ hold, the above spectral
indices are approximately given by
\begin{eqnarray}
& &n_{\cal R}-1= -2(2\epsilon_1+\epsilon_2-\epsilon_3+\epsilon_4)\,, \\
& &n_{T}=-2(\epsilon_1+\epsilon_6)\,.
\end{eqnarray}
Since $\nu_{\cal R} \simeq 3/2 \simeq \nu_{T}$ in this case,
the tensor-to-scalar ratio (\ref{ratiodef}) reads
\begin{eqnarray}
\label{ratio2}
r=8 \frac{\omega \dot{\phi}^2+\frac{3(\dot{F}+Q_1)^2}
{2F+Q_2}+Q_3}{\left(H+\frac{\dot{F}+Q_1}{2F+Q_2}
\right)^2 \left(F+\frac{Q_2}{2}\right)}
\left(\frac{c_{\cal R}}{c_T} \right)^3\,.
\end{eqnarray}

Let us consider the case in which the GB term is absent ($c_1=0$).
Since $Q_T=F$ and $\epsilon_6=\epsilon_3$, one has
$n_T=-2(\epsilon_1+\epsilon_3)$.
{}From Eq.~(\ref{f2}) we obtain the relation
$\omega \dot{\phi}^2/H^2F-2c_2 \xi \dot{\phi}^4/H^2F
\simeq 2(\epsilon_1+\epsilon_3)$.
Then with the use of Eq.~(\ref{cRdef}) the tensor-to-scalar ratio
(\ref{ratio2}) is simply given by
$r=16(\epsilon_1+\epsilon_3)c_{\cal R}$.
Hence we find
\begin{eqnarray}
\label{consistency}
\label{rac1=0}
r=-8c_{\cal R}n_{T} \quad ({\rm for}~~c_1=0).
\end{eqnarray}

When the $c_2 (\nabla \phi)^4$ term is absent,
one has $c_{\cal R}=1$, thereby reducing to the standard
consistency relation: $r=-8n_T$ \cite{BTW}.
Provided that ${\cal L}_c=0$, this standard consistency relation
holds even for scalar-tensor models
characterized by the action
(\ref{action}) \cite{Gum,Hwang05}.

\section{Einstein frame}
\label{einstein}

The action (\ref{action}) is transformed to the one in the Einstein
frame by a conformal transformation \cite{Maeda}:
\begin{eqnarray}
\label{conformal}
\hat{g}_{\mu \nu}=\Omega g_{\mu \nu}\,,\quad
\Omega=F\,,
\end{eqnarray}
where a hat represents quantities in the Einstein frame.
For later convenience, we write the correction term
${\cal L}_{c}$ as ${\cal L}_c={\cal L}_{\rm GB}
-(1/2)c_2 \xi(\phi) (\nabla \phi)^4$, where ${\cal L}_{\rm GB}$
is the contribution of the GB term.
Then the action in the Einstein frame is
\begin{eqnarray}
\label{eaction}
S_{E} &=& \int {\rm d}^4 \hat{x} \sqrt{-\hat{g}}
\Biggl[ \frac{\hat{R}}{2}
+K(\phi) X+ L(\phi) X^2 \nn
&& \hspace{20mm}
-\hat{V} (\phi)+{\cal \hat{L}}_{\rm GB}
\Biggr]\,,
\end{eqnarray}
where $X=-(1/2)(\hat{\nabla} \phi)^2
=(1/2)(\rd \phi/\rd \hat{t})^2$ and
\begin{eqnarray}
\label{Kphi}
& & K(\phi)=\frac32 \left( \frac{F_{,\phi}}
{F} \right)^2+\frac{\omega}{F}\,, \\
& & L(\phi)=-2c_2 \xi(\phi)\,, \\
\label{VEdef}
& & \hat{V} (\phi)=\frac{V(\phi)}{F^2}\,.
\end{eqnarray}

Let us introduce a perturbed metric in the Einstein frame:
\begin{eqnarray}
\rd \hat{s}^2 &=& \Omega \rd s^2 \nonumber \\
&=&-(1+2\hat{A})\rd \hat{t}^2 +2\hat{a}
\partial_{i} \hat{B}
\rd \hat{x}^i \rd \hat{t} \nonumber \\
& &+\hat{a}^2 \left[(1+2\hat{\psi})\delta_{ij}
+2\partial_{ij} \hat{E}+2\hat{h}_{ij}
\right]\rd \hat{x}^i \rd \hat{x}^j\,.~~~
\label{pmetricEin}
\end{eqnarray}
We decompose the conformal factor into the background and
the perturbed part as
\begin{eqnarray}
\Omega({\bf x}, t)=\bar{\Omega}(t)
\left(1+\frac{\delta \Omega({\bf x}, t)}
{\bar{\Omega}(t)}\right)\,.
\label{Omega}
\end{eqnarray}
Then the following relations are derived:
\begin{eqnarray}
& &\hat{a}=a\sqrt{\Omega},~~\rd \hat{t}=
\sqrt{\Omega}\rd t, ~~
\hat{H}=\frac{1}{\sqrt{\Omega}}
\left(H+\frac{\dot{\Omega}}{2\Omega}\right),
\nonumber \\
& &\hat{A}=A+\frac{\delta \Omega}{2\Omega},~~
\hat{\psi}=\psi+\frac{\delta \Omega}
{2\Omega},
\label{relation}
\end{eqnarray}
where a ``bar'' is dropped from the expression of
$\bar{\Omega}(t)$.

{}From these relations, one can show that the
curvature perturbation in the Einstein frame
coincides with that in the Jordan frame:
\begin{eqnarray}
\hat{\cal R} &\equiv& \hat{\psi}-\frac{\hat{H}}
{{\rm d}\hat{\phi}/{\rm d}\hat{t}}\delta \hat{\phi}
\nonumber \\
&=& \psi-\frac{H}{\dot{\phi}}\delta\phi
={\cal R}\,.
\label{calRein}
\end{eqnarray}
Since tensor perturbations are also invariant
under a conformal transformation,
the power spectra of scalar and tensor
perturbations satisfy
\begin{eqnarray}
\hat{\cal P}_{\cal R}={\cal P}_{\cal R}\,,~~~~
\hat{\cal P}_{T}={\cal P}_{T}\,.
\label{RT}
\end{eqnarray}
We also note that the comoving wavenumber, $k^2$, is
invariant under a conformal transformation  (since
in the Fourier space $\Delta=-k^2$ is invariant).
Hence the amplitudes and the spectral indices of
power spectra based upon $\cal R$ and $h_{ij}$
in the string frame coincide with those in the Einstein frame.

\subsection{Models without the GB term ($c_1=0$)}

In this section, we investigate a situation
in which the contribution of the GB term is
absent [${\cal \hat{L}}_{\rm GB}=0$ in
Eq.~(\ref{eaction})].
In order to derive $n_{\cal R}$ and other physical quantities for the action
(\ref{eaction}) without the GB term, it is sufficient to
use the formula obtained in the previous section
by replacing $F(\phi) \to 1$, $\omega(\phi) \to K(\phi)$,
$c_1 \to 0$ and $c_2 \xi(\phi) \to -(1/2)L(\phi)$.
Introducing the following parameters
\begin{eqnarray}
\hat{\epsilon}_{1}=-\frac{\rd \hat{H}/\rd \hat{t}}
{\hat{H}^2}\,, \quad
\hat{\epsilon}_{2}=\frac{\rd^2 \phi/\rd \hat{t}^2}
{\hat{H}(\rd \phi /\rd \hat{t})}\,, \quad
\hat{\epsilon}_{4}=\frac{\rd \hat{E}/\rd \hat{t}}
{2\hat{H}\hat{E}}\,,
\label{epein}
\end{eqnarray}
where $E=K+6LX$, the spectral index $n_{\cal R}$
of curvature perturbations is
\begin{eqnarray}
\label{nR1}
n_{\cal R}-1 &=& 3-\left| \frac{3+\hat{\epsilon}_1+2\hat{\epsilon}_2+
2\hat{\epsilon}_4}{1-\hat{\epsilon}_1} \right| \\
\label{nR2}
& \simeq & -2(2\hat{\epsilon}_1+\hat{\epsilon}_2+\hat{\epsilon}_4)\,.
\end{eqnarray}
The first equality is valid under the condition that the derivative
terms of $\epsilon_i$ vanish or they are negligible compared to
other terms, whereas the second
approximate equality is valid under the
slow-roll approximation.
Similarly the spectral index of tensor perturbations is
\begin{eqnarray}
\label{nT1}
n_T &=& 3-\left| \frac{3-\hat{\epsilon}_1}
{1-\hat{\epsilon}_1} \right| \\
\label{nT2}
&\simeq & -2\hat{\epsilon}_1\,.
\end{eqnarray}
When $|\epsilon_i| \ll 1$ the tensor-to-scalar ratio
$r$ satisfies the consistency relation (\ref{consistency})
with $c_{\cal R}$ given by \cite{Garriga}
\begin{eqnarray}
\label{cRki}
c_{\cal R}^2=\frac{K+2LX}{K+6LX}\,.
\end{eqnarray}
\subsection{Models without the correction term
($c_1= c_2=0$)
}

Let us next consider the case in which the correction term
${\cal L}_c$ is absent ($c_1=c_2=0$).
When $K(\phi)$ is positive, we introduce
a new scalar field $\varphi$ as
\begin{eqnarray}
\label{vphidef}
\varphi=\int \sqrt{K(\phi)}\,\rd \phi\,.
\end{eqnarray}
Then the action (\ref{eaction}) can be written in
a canonical form
\begin{eqnarray}
\label{eaction2}
S_{E} = \int {\rm d}^4 \hat{x} \sqrt{-\hat{g}}
\left[ \frac{\hat{R}}{2}-\frac12 (\hat{\nabla} \varphi)^2
-\hat{V}(\varphi (\phi)) \right]\,.
\end{eqnarray}
If $K(\phi)$ is negative, we just need to define
$\varphi=\int \sqrt{-K(\phi)}\,\rd \phi$ with the
change of sign of the kinetic term in
Eq.~(\ref{eaction2}).

Introducing the parameters
\begin{eqnarray}
\hat{\epsilon}_{1}=-\frac{\rd \hat{H}
/\rd \hat{t}}{\hat{H}^2}\,, \quad
\hat{\epsilon}_{2}=\frac{\rd^2 \varphi/\rd \hat{t}^2}
{\hat{H}(\rd \varphi/\rd \hat{t})}\,,
\end{eqnarray}
the spectral indices $n_{\cal R}$ and $n_{T}$ are
given by Eqs.~(\ref{nR1})-(\ref{nT2}) with $\hat{\epsilon}_{4}=0$.
{}From Eq.~(\ref{ratiodef}), the tensor-to-scalar ratio $r$ is
\begin{eqnarray}
\label{ratioda}
r=16\hat{\epsilon}_1 \left(
\frac{\Gamma (\hat{\nu}_{T})}
{\Gamma (\hat{\nu}_{\cal R})}
\right)^2\,,
\end{eqnarray}
where we have used $\rd \hat{H} /\rd \hat{t}=
-(\rd \varphi/\rd \hat{t})^2/2$.
Note that this relation is obtained without using
the slow-roll approximation.
In order to satisfy the condition $r \ll 1$, which is
supported from observations,
we should have $\hat{\epsilon}_1 \ll 1$
provided that $\Gamma (\hat{\nu}_{T})
/\Gamma(\hat{\nu}_{\cal R})$ is of order 1.
The condition to obtain a nearly scale-invariant scalar
perturbation ($n_{\cal R} \simeq 1$) then gives two
cases: (i) $\hat{\epsilon}_2 \simeq 0$ or
(ii) $\hat{\epsilon}_2 \simeq -3$.
The former corresponds to the standard slow-roll inflation.
The latter corresponds to a constant scalar-field potential
($\hat{V}_{,\varphi}=0$).
In both cases, we have $r \simeq 16 \epsilon_1$
since $\nu_{\cal R} \simeq \nu_T \simeq 3/2$.

One may define standard slow-roll parameters
in terms of the slope of the potential
\begin{eqnarray}
\label{epeta}
\hat{\epsilon}=\frac{1}{2} \left(
\frac{\hat{V}_{,\varphi}}{\hat{V}} \right)^2\,, \quad
\hat{\eta}=\frac{\hat{V}_{,\varphi \varphi}}
{\hat{V}}\,.
\end{eqnarray}
Under a slow-roll approximation ($|\epsilon_{i}| \ll 1$) one has
$\hat{\epsilon}_1 \simeq  \hat{\epsilon}$ and
$\hat{\epsilon}_2 \simeq \hat{\epsilon}-\hat{\eta}$.
Hence we get the following usual
formulae \cite{Lidsey97,BTW}:
\begin{eqnarray}
n_{\cal R}-1 \simeq -6\hat{\epsilon}+2\hat{\eta}\,, \quad
n_{T} \simeq -2\hat{\epsilon} \,, \quad
r \simeq 16\hat{\epsilon}\,.
\end{eqnarray}
%

\section{Dilaton gravity without the correction term (${\cal L}_c=0$)}
\label{dilaton}

Let us first consider dilaton gravity without the correction
term ${\cal L}_{c}$ and study the possibility
to obtain scale-invariant cosmological perturbations.
The low-energy string effective action corresponds to
$F(\phi)=e^{-\phi}$, $\omega(\phi)=-e^{-\phi}$,
$V(\phi)=0$ at the tree level.
In the Einstein frame the system is described by
a minimally coupled field $\varphi$ without a potential,
thus giving the following solution \cite{PBB}:
\begin{eqnarray}
\label{evolution}
\hat{a} \propto |\hat{t}|^{1/3}\,,\quad
\hat{H}=\frac{1}{3\hat{t}}\,,\quad
\left(\frac{\rd \varphi}{\rd \hat{t}}
\right)^2=\frac{2}{3\hat{t}^2}\,.
\end{eqnarray}
When $\hat{t}<0$, this solution corresponds to a collapsing universe.
In the Pre-Big-Bang cosmology, the universe contracts in the Einstein frame
for $\hat{t}<0$, which is followed by the bounce around
$\hat{t}=0$ because of the effect of higher-order
loop and derivative corrections \cite{higherpapers,Foffa}.
Note that a causal mechanism for perturbations works in the
collapsing universe as in the case of standard inflation.
If we consider a growing field $\varphi$ for $\hat{t}<0$,
we have $\rd \varphi/\rd \hat{t}=-\sqrt{2/3}(1/\hat{t})$ .

We find from Eq.~(\ref{evolution}) that $\hat{\epsilon}_{1}$
and $\hat{\epsilon}_2$ defined in Eq.~(\ref{epein})
are constants, i.e.,
\begin{eqnarray}
\hat{\epsilon}_1=3\,, \quad
\hat{\epsilon}_2=-3\,.
\end{eqnarray}
In this case one can use the formula (\ref{nR1}), (\ref{nT1})
and (\ref{ratioda}) with $\hat{\epsilon}_{4}=0$, which gives
\begin{eqnarray}
\label{PBBspe}
n_{\cal R}=4\,, \quad n_T=3\,, \quad
r=48\,.
\end{eqnarray}
This is a highly blue-tilted spectrum and a large tensor-to-scalar
ratio incompatible with observations.
Although these perturbations correspond to the ones which
are generated before the bounce, it was shown that these
spectra are preserved even long after the
bounce if $\alpha'$ curvature and derivative corrections are
taken into account \cite{TBF02}.

The reason why the spectrum is highly blue-tilted is that
the system is dominated by the kinetic energy of the
scalar field. In order to obtain nearly scale-invariant
spectra, we should require that $\hat{\epsilon}_{1}$ and
$\hat{\epsilon}_{2}$ are much smaller than 1.
This situation is not realized unless a slowly
varying potential is present in the Einstein frame.
In order to investigate the possibility of obtaining
nearly scale-invariant perturbation spectra,
in what follows, we study models with
general forms of $F(\phi)$ and $\omega(\phi)$
together with the dilaton potential $V(\phi)$.
Note that the results given in Eq.~(\ref{PBBspe}) for the Pre-Big-Bang
model were already derived in past
works (e.g.~Ref.~\cite{Brustein94}), but the results we present
below in more general models are new.

It follows from Eq.~(\ref{ratioda}) that we must have
$\hat{\epsilon} \ll 1$ to realize the condition $r \ll 1$ unless the
$(\Gamma (\hat{\nu}_T)/\Gamma (\hat{\nu}_{\cal R}))^2$ term
is much smaller than unity.
Let us clarify whether or not this condition
can be satisfied in dilaton gravity. For the coupling
$F(\phi)=e^{-\phi}$ and $\omega(\phi)=-e^{-\phi}$,
the dilaton potential in the Einstein frame is given by
\begin{eqnarray}
\hat{V}=e^{2\phi} V(\phi)=
e^{2\sqrt{2} \varphi} V(\sqrt{2}\varphi)\,,
\end{eqnarray}
where we have used Eqs.~(\ref{VEdef}) and (\ref{vphidef}).
Since $\varphi=\phi/\sqrt{2}$, the parameter
$\hat{\epsilon}$ defined in Eq.~(\ref{epeta}) is given by
\begin{eqnarray}
\hat{\epsilon}
=\left( 2+ \frac{V_{,\phi}}{V} \right)^2\,.
\end{eqnarray}
Even when a slowly varying potential is present in the string frame
($|V_{,\phi}/V| \ll 1$), we get $\hat{\epsilon} \simeq 4$, thus
leading to a large tensor-to-scalar ratio.
The potential which gives $\hat{\epsilon}=0$ is characterized by
\begin{eqnarray}
\label{postring}
V(\phi)=V_0 e^{-2 \phi}\,,
\end{eqnarray}
where $V_{0}$ is a constant.
In fact in this case the potential in the Einstein frame is exactly
constant, thus giving $n_{\cal R}=1$ and $r=0$
because $\hat{\epsilon}_1=0$ and $\hat{\epsilon}_2=-3$.
Hence for the coupling $F(\phi)=e^{-\phi}$ and
$\omega(\phi)=-e^{-\phi}$, it is required that the dilaton
potential in the string frame (approximately)
takes the form (\ref{postring})
to realize $n_{\cal R} \simeq 1$ and $r \ll 1$.
We note however that the potential of the dilaton is absent
in the perturbative regime we considered above.

Let us consider more general functions of
$F(\phi)$ and $\omega (\phi)$ with a dilaton potential $V(\phi)$.
Then the slow-roll parameter $\hat{\epsilon}$ is given by
\begin{eqnarray}
\hat{\epsilon}=\frac{1}{3F_{,\phi}^2/F^2+2\omega/F}
\left(\frac{V_{,\phi}}{V}-2\frac{F_{,\phi}}
{F}\right)^2\,.
\end{eqnarray}
This shows that there are three cases in which
the condition $\hat{\epsilon} \ll 1$ is satisfied:

(i) $V_{,\phi}/V \simeq 2F_{,\phi}/F$,

(ii) $1 \gg |V_{,\phi}/V| \gg |2F_{,\phi}/F|$,

(iii) $1 \gg |2F_{,\phi}/F| \gg |V_{,\phi}/V|$.

Note that we are considering the situation where the
$|\omega/F|$ term is of order unity.
In a special case with $|\omega/F| \gg 1$, the condition
$\hat{\epsilon} \ll 1$ can be satisfied even for
$|\omega/F| \gg (V_{,\phi}/V)^2 \gg (F_{,\phi}/F)^2 \gg 1$.
One may also think that the slow-roll condition can be fulfilled
even for $V=0$ provided $|F_{,\phi}/F| \ll 1$, but
the definition of $\hat{\epsilon}$ itself is not meaningful
in this case because of the absence of the potential.

The case (i) is equivalent to the condition
\begin{eqnarray}
\label{VF}
V(\phi) \propto F^2(\phi)\,,
\end{eqnarray}
which, in the Einstein frame, corresponds to
a cosmological constant from Eq.~(\ref{VEdef}).

The cases (ii) and (iii) are difficult to realize if the dilaton coupling
$F(\phi)$ changes rapidly as in the tree-level action.
For the coupling $F(\phi) \sim e^{\lambda \phi}$
we require the condition $|\lambda| \ll 1$, but
it is generally difficult to take a control of the rapid
evolution of the dilaton in the non-perturbative regime.
However one may consider so-called a runaway dilaton
scenario \cite{runaway}
in which the field is gradually decoupled from gravity
in the regime $e^{\phi} \gg 1$ (after the field
passes the non-perturbative region).
In this scenario the functions $F(\phi)$ and $\omega(\phi)$
are assumed to take the following forms for $e^{\phi} \gg 1$:
\begin{eqnarray}
& &F(\phi)=C_1+D_1 e^{-\phi}+{\cal O} (e^{-2\phi})\,,\\
& & \omega (\phi)=C_2+D_2e^{-\phi}+{\cal O} (e^{-2\phi})\,.
\end{eqnarray}
Let us study the case in which the potential of the dilaton
is present in this region.
Then the slow-roll parameter $\hat{\epsilon}$ is
approximately given by
\begin{eqnarray}
\label{ep2}
\hat{\epsilon} \simeq \frac{C_1}{2C_2} \left(
\frac{V_{,\phi}}{V}+2\frac{D_1}{C_1}
e^{-\phi} \right)^2\,.
\end{eqnarray}
In the cases (ii) and (iii) we have $|V_{,\phi}/V| \gg |2(D_1/C_1)e^{-\phi}|$
and $|V_{,\phi}/V| \ll |2(D_1/C_1)e^{-\phi}|$, respectively.
One may consider a potential in which the field has a minimum
at $\phi=\phi_0$, i.e., $V(\phi)=\lambda_n (\phi-\phi_0)^n$.
When $1 \gg |V_{,\phi}/V| \gg |2(D_1/C_1)e^{-\phi}|$, we obtain
$n_{\cal R}$ and $r$ as in the case of standard chaotic
inflation \cite{BTW}. In other words, it is necessary that the dilaton
is effectively decoupled from gravity in order to realize
standard slow-roll inflation in the Einstein frame.
Note that inflation is also realized for
$|V_{,\phi}/V| \ll |2(D_1/C_1)e^{-\phi}| \ll 1$ except for the
case in which the potential of the dilaton vanishes.

The above discussion shows how the coupling of dilaton to gravity alters
the spectra of density perturbations. Except for the specific case
given in Eq.~(\ref{VF}), the dilaton needs to be almost
decoupled from gravity before perturbations on cosmologically
relevant scales are generated.

\section{Kinetic inflation ($c_1=0, c_2 \neq 0, V=0$)}
\label{kinetic}

When $c_{1}=0$ and $c_{2} \neq 0$,
it is known that kinetic inflation can be realized in
the Einstein frame even in the absence of the
dilaton potential \cite{kinf}.
This corresponds to the action (\ref{eaction})
with $\hat{V}=0$ and $\hat{{\cal L}}_{\rm GB}=0$.
Then the background equations in the Einstein frame are given by
\begin{eqnarray}
\label{baki1}
& & 3\hat{H}^2=\hat{\rho} \,, \\
\label{baki2}
& & 2\dot{\hat{H}}=-(\hat{\rho}+\hat{p})\,,
\end{eqnarray}
where
\begin{eqnarray}
\label{prho}
& & \hat{p}=K(\phi)X+L(\phi)X^2\,, \\
& & \hat{\rho}=K(\phi)X+3L(\phi)X^2\,.
\end{eqnarray}

In the weak coupling regime characterized by
$e^{\phi} \ll 1$, one has $F(\phi)=e^{-\phi}$,
$\omega (\phi)=-e^{-\phi}$, $\xi(\phi)=\lambda e^{-\phi}$
and $c_2=-1$. Hence the functions $K(\phi)$ and $L(\phi)$
are given by $K(\phi)=1/2$ and $L(\phi)=2\lambda e^{-\phi}$.
Since $\lambda=-1/4, -1/8$ for bosonic and heterotic strings,
respectively, the function $L(\phi)$ is negative at the tree level.
In the strong coupling regime ($e^{\phi} \gtrsim 1$),
the forms of the functions $K(\phi)$ and $L(\phi)$
are expected to be modified by higher-order quantum effects.

{}From the above equations, we find that a de Sitter solution
($\dot{\hat{H}}=0$) is present when the condition
\begin{eqnarray}
\label{KL}
K+2LX=0\,,
\end{eqnarray}
is satisfied.
In this case one gets $3\hat{H}^2=K^2/(4L)$, thereby requiring
$L>0$. Then from Eq.~(\ref{KL}), the function $K$
is negative for the existence of the de Sitter solution.
In what follows, we concentrate on the case
where $K<0$ and $L>0$.
Note that ghost condensate model proposed in Ref.~\cite{Arkani}
corresponds to $K=-1$ and $L=1$.

While $K(\phi)$ is positive at the tree level, it can change sign
in the strongly coupled regime.
In fact taking into account two-derivative perturbative loop corrections
to the Kahler potential derived from heterotic string theory
compactified on a $Z_N$ orbifold, the function $\omega(\phi)$
is subject to change \cite{Foffa}
\begin{eqnarray}
\omega (\phi)=-e^{-\phi} \left[ 1+\frac32
\frac{be^{\phi} (6+be^{\phi})}{(3+be^{\phi})^2}
\right]\,,
\end{eqnarray}
where $b$ is a positive constant of order unity.
Then from Eq.~(\ref{Kphi}) with $F(\phi)=e^{-\phi}$,
the function $K(\phi)$ is given by
\begin{eqnarray}
\label{kcon}
K(\phi)=\frac12-\frac32
\frac{be^{\phi} (6+be^{\phi})}{(3+be^{\phi})^2}\,,
\end{eqnarray}
which is negative for $be^{\phi}>3(\sqrt{6}-2)/2$.
This function has the asymptotic behaviors
$K(\phi) \simeq -1+27e^{-2\phi}/2b^2 \to -1$ for
$e^{\phi} \gg 1$ and $K(\phi) \simeq 1/2-be^{\phi} \to 1/2$ for
$e^{\phi} \ll 1$.
Note that the above corrections are obtained perturbatively,
so they may not be valid in a full non-perturbative regime.
But this example is instructive to show the possibility to
obtain negative values of $K(\phi)$ after the system enters
the strongly coupled region.

When $K(\phi)$ is negative, it is possible to realize kinetic
inflation provided that $L(\phi)>0$.
At the tree level, $L(\phi)$ is negative for bosonic and heterotic strings,
but it can be positive in the strongly coupled region.
One may write $\xi(\phi)$ in terms of the sums of the tree-level and
$n$-loop corrections in the form
\begin{eqnarray}
\label{xiphi}
\xi (\phi)=\sum_{n=0} C_{n} e^{(n-1)\phi}\,,
\end{eqnarray}
where $n=0$ corresponds to the tree-level term
with $C_0=1$. Unfortunately all the coefficients $C_n$
have not been determined so far.
So in what follows, we discuss the theory with the
couplings given in (\ref{xiphi}) with various $C_n$.
This is at least useful to clarify the condition under which
kinetic inflation generates nearly scale-invariant
density perturbations.

It follows from Eq.~(\ref{cRki}) that $c_{\cal R}^2 \simeq 0$
along the de Sitter point (\ref{KL}) and
$c_{T}^2=1$ from Eq.~(\ref{cT2}).
By using the formula given in Eqs.~(\ref{nR2}), (\ref{nT2}) and
(\ref{consistency}), we obtain
\begin{eqnarray}
& & n_{\cal R}-1=-2 \Biggl[ \frac{6(K+2LX)}{K+3LX}
+\frac{\dot{X}}{2\hat{H}X} \nonumber \\
& & \hspace{22mm} +
\frac{(K_{,\phi}+6XL_{,\phi})\dot{\phi}+6L\dot{X}}
{2\hat{H} (K+6LX)} \Biggr]\,, \\
& &  n_{T}=-\frac{6(K+2LX)}{K+3LX}\,, \\
& &  r=-8 \sqrt{\frac{K+2LX}{K+6LX}}\,n_{T}\,.
\end{eqnarray}
Then along the de Sitter fixed point ($X \simeq -K/2L$), one has
\begin{eqnarray}
n_T \simeq 0\,,\quad
r \simeq 0\,.
\end{eqnarray}
Meanwhile the spectrum of curvature perturbations
is given by
\begin{eqnarray}
n_{\cal R}-1=\left( -\frac{2K_{,\phi}}{\hat{H}K}+
\frac{L_{,\phi}}{\hat{H}L} \right) \dot{\phi}\,.
\end{eqnarray}
By using $\hat{H}=\sqrt{-K/12} |\dot{\phi}|$ along the de Sitter
solution, this is written as
\begin{eqnarray}
\label{speRT}
n_{\cal R}-1=\frac{{\rm sign}(\dot{\phi})}{\sqrt{-K/12}}
\left(-\frac{2K_{,\phi}}{K}+\frac{L_{,\phi}}{L} \right)\,.
\end{eqnarray}

Let us first investigate the case in which $L(\phi)$ is a constant.
We are interested in the evolution of dilaton from a
strongly coupled region ($e^{\phi} \gg 1$ and $K<0$) to
a weakly coupled region ($e^{\phi} \ll 1$ and $K>0$),
which means that $\dot{\phi}<0$ and $K_{,\phi}<0$.
Alternatively it is possible to consider the case with $\dot{\phi}>0$
and $K_{,\phi}>0$ in which the system gradually decouples from
gravity in the region $e^{\phi} \gg 1$
as in the runaway dilaton scenario.
In both cases, we obtain a blue-tilted spectrum $n_{\cal R}>1$.

We next study a situation in which one of the terms in
Eq.~(\ref{xiphi}), say $e^{\lambda \phi}$, dominates over
other terms in the strongly coupled region (as in the dilatonic
ghost condensate model proposed in Ref.~\cite{Piazza}).
This then gives
\begin{eqnarray}
n_{\cal R}-1=\frac{{\rm sign}(\dot{\phi})}{\sqrt{-K/12}}
\left(\lambda -\frac{2K_{,\phi}}{K} \right)\,.
\end{eqnarray}
If the terms higher than the one-loop corrections are dominant,
i.e., $\lambda \ge 1$, the power spectrum strongly departs from
the scale-invariant one.
Thus in order to be compatible with observations,
we must require that
the function $\xi(\phi)$ be nearly constant
in the regime $e^{\phi} \gg 1$.
Taking the terms up to the first order in Eq.~(\ref{xiphi}), we have
$L_{,\phi}/L \simeq -e^{-\phi}/C_1>0$ (because $C_1$ is
negative for the positivity of $L$).
If $L_{,\phi}/L$ is larger than $2K_{,\phi}/K$, the
red-tilted spectrum follows for $\dot{\phi}<0$.
For example, if the function $K(\phi)$ is given
by Eq.~(\ref{kcon}), we find
\begin{eqnarray}
n_{\cal R}-1 \simeq
\frac{{\rm sign}(\dot{\phi})}{\sqrt{-K/12}}
\left(-\frac{54}{b^2}
e^{-2\phi}+\frac{1}{|C_1|} e^{-\phi}  \right)\,,
\end{eqnarray}
where the second term in the square bracket dominates over the
first one. Hence in this case one gets a red-tilted spectrum for
$\dot{\phi}<0$. Of course if the function $K(\phi)$ takes a
different form such as $K(\phi)=-{\rm tanh}\,(\lambda \phi)$,
it can happen that the spectrum is blue-tilted.

Thus the spectral index of curvature perturbations
depends upon the forms of $K(\phi)$ and $L(\phi)$
in the case of de Sitter solutions.
Let us estimate $n_{\cal R}$ on the cosmologically
relevant scales observed in Cosmic Microwave Background
anisotropies. For simplicity, we choose the function $K(\phi)$ given in
Eq.~(\ref{kcon}) with constant $L(\phi)$.
Along the de Sitter solution (\ref{KL}) one has
$\dot{\phi}=\sigma \sqrt{-K/L}$ and
$\hat{H}=-K/2\sqrt{3L}$ where $\sigma=\pm 1$.
This then gives the number of $e$-foldings:
\begin{eqnarray}
N \equiv \int_{t}^{t_f} \hat{H} {\rm d}\hat{t}=
\frac{\sigma}{2\sqrt{3}} \int_{\phi}^{\phi_f}
\sqrt{-K} {\rm d}\phi\,,
\end{eqnarray}
where the subscript ``{\it f}'' represents the values at the end of inflation.
If inflation occurs in the region $be^{\phi} \gg 1$,
we get $N \simeq (\phi-\phi_f)/2\sqrt{3}$
(here negative $\sigma$ is chosen).

Considering homogeneous perturbations $\delta X$ around
$X$ \cite{kinf}, we find that $p_{,X}=2L \delta X$
and $\dot{\rho}=-2\sqrt{3\rho}X p_{,X}$. Then one has
$\delta X/X =-(\sigma /\sqrt{3})(-K)_{,\phi}/(-K)^{3/2}$.
Inflation ends when the $\delta X/X$ term grows of order unity.
Taking the criterion $(-K)_{,\phi}/(-K)^{3/2} =\sqrt{3}$ for the end of
inflation, we obtain $be^{\phi_f} \sim 1.75$ for the model (\ref{kcon}).
This gives $be^{\phi}=1.75 e^{2\sqrt{3}N}$.
Then we finally get the spectral index
in terms of the function of $N$:
\begin{eqnarray}
n_{\cal R}-1=4\sqrt{3} \frac{(-K)_{,\phi}}
{(-K)^{3/2}} \simeq 60 e^{-4\sqrt{3}N}\,.
\end{eqnarray}
For the cosmologically relevant scales ($N \sim 60$),
the spectrum is extremely close to scale-invariant one.
This situation does not change
provided that $K(\phi)$ and $L(\phi)$ are nearly constants
with an exponentially suppressed factor.
These models, which give $n_{\cal R} \simeq 1$ and $r \simeq 0$,
are interesting to confront with future
high-precision observations, since recent WMAP3 data
do not favor an exact scale-invariant scalar perturbation
with a vanishing tensor-to-scalar ratio \cite{WMAP3}.

We note that there are some cases in which the spectrum is tilted
from the scale-invariant one.
As an example, let us consider the realization of power-law
kinetic inflation which is characterized by
\begin{eqnarray}
\label{plaw}
\hat{a} \propto \hat{t}^{1/\gamma}\,, \quad
\hat{H}=\frac{1}{\gamma \hat{t}}\,,
\end{eqnarray}
where $0<\gamma<1$.
For the choice $L(\phi)=-K(\phi)$, we find from
Eqs.~(\ref{baki1}) and (\ref{baki2}) that
\begin{eqnarray}
X=\frac{3-\gamma}{3(2-\gamma)}\,, \quad
K(\phi)=-\frac{6(2-\gamma)}{\gamma^2 (\phi -\phi_0)^2}\,,
\end{eqnarray}
where $\phi_0$ is the initial value of the field.
Hence from Eqs.~(\ref{nR1}), (\ref{nT1}) and (\ref{ratiodef})
with $\hat{\epsilon}_1=-\hat{\epsilon}_4=\gamma$,
$\hat{\epsilon}_2=0$ and $c_{\cal R}^2=
\gamma/(3(4-\gamma))>0$, we obtain
\begin{eqnarray}
\label{const1}
& &n_{\cal R}-1=n_T=-\frac{2\gamma}{1-\gamma}\,,\\
&& r = 16 \gamma \left[\frac{\gamma}{3(4-\gamma)}
\right]^{\frac{1+\gamma}{2(1-\gamma)}}\,.
\label{const2}
\end{eqnarray}
This indicates that the spectra of scalar and tensor perturbations
are red-tilted ($n_{\cal R}<1$ and $n_{T}<0$).
The parameter $\gamma$ is a measure of
the departure from the Harrison-Zeldvich scale-invariant
spectra ($n_{\cal R}-1=n_T=0$ and $r=0$).
The recent WMAP3 data \cite{WMAP3} give the constraints
$n_{\cal R}=0.987^{+0.019}_{-0.037}$ and $r<0.55$
at the $2\sigma$ level for the $\Lambda$CDM
model without the running of scalar perturbations.
The constraint~(\ref{const1}) then requires that $\gamma$ should be
$\gamma<0.024$. This in turn gives $r<0.0148$ which is
much smaller than 0.55.
This shows that the constraint from $n_{\cal R}$ gives a severer
bound on $\gamma$.
Finally we note that in order to end the power-law inflation
the functions $K(\phi)$ and $L(\phi)$ need to be modified
in a suitable way.

\section{Models with the correction term}
\label{correctionsec}

In this section we investigate models in which the correction term
${\cal L}_c$ is present. We are interested in the realization of
inflation without using a dilaton potential.\footnote{Recently there
are a number of works which aim to explain late-time acceleration
with the presence of the potential $V(\phi)$ \cite{GBdenergy}.
For example, in the case of an exponential potential,
a de Sitter fixed point appears because of the effect
of the GB term \cite{Koi}.}
Basically one may consider the following two situations:
(i) $F(\phi)$ asymptotically approaches a constant
value (as in the runaway dilaton scenario), or
(ii) $F(\phi)$ is the sum of exponential terms.
In the case (i), we set $F(\phi)=1$ by assuming that
the time-derivatives of $F(\phi)$ in Eqs.~(\ref{f1}) and
(\ref{f2}) are negligible relative to others.
In the case (ii), we study models in which
one of the exponential terms dominates over
others, i.e., $F(\phi)=e^{\mu \phi}$.

\subsection{Models with $F(\phi) \sim$ constant}

In order to understand the effect of the GB term
to realize inflation, we first study the case
$c_{2}=0$ and $F(\phi)=1$.
Then Eqs.~(\ref{f1}) and (\ref{f2}) yield
\begin{eqnarray}
&& 6H^2 = \omega \dot \phi^2+24c_1 \dot{\xi}H^3\,,
\label{s1} \\
&& 2\dot H = -\omega \dot \phi^2 +4c_1
\left[ H^2 \ddot{\xi}+(2H\dot{H}-H^3)\dot{\xi}
\right]\,.
\label{s2}
\end{eqnarray}
We search for power-law solutions given by
\begin{eqnarray}
\label{hp}
a \propto t^{1/\gamma}\,, \quad
H=\frac{1}{\gamma t}\,.
\end{eqnarray}
At the end of this subsection we also discuss
the case of de Sitter solutions ($H={\rm const.}$).
Eliminating the $\omega \dot \phi^2$ term from
Eqs.~\p{s1} and \p{s2}, we find that in order to
get the solution~\p{hp}, we must have
\ba
c_1 \dot \xi = \alpha t\,,
\label{24}
\ea
where $\alpha$ is a constant.
To realize a positive energy density $\rho_c$ for
$t>0$ we require $\alpha$ is positive, which
is assumed hereafter.
The relation between $\alpha$ and $\gamma$ is given by
\ba
\gamma= \frac{2\alpha+3 \pm \sqrt{4\alpha^2-28\alpha+9}}
{2}\,.
\label{sol2}
\ea

An inflationary solution is obtained for $\gamma<1$.
We thus choose the lower sign in \p{sol2} and get the constraint
on $\alpha$:
\ba
0< \alpha < 1/4\,.
\ea
It follows from Eqs.~\p{s1}, \p{24} and \p{sol2} that
\ba
\rho_c=\dfrac{6(3-\gamma)}{5-\gamma}H^2>3H^2\,.
\ea
This implies that in order for Eq.~\p{s1} to have an inflationary
solution we must have $\omega<0$, i.e., a phantom-type scalar field.
For simplicity, we choose $\omega=-1$.
But this does not necessarily mean that the weak energy
condition is violated when the correction term ${\cal L}_c$
is present. In fact from Eq.~(\ref{s2}) or (\ref{hp}), we find
$\dot{H}= - 1/(\gamma t^2)<0$ with $0<\gamma<1$.
Taking the positive sign of $\dot{\phi}$, we obtain
\ba
& & \phi=\frac{1}{\gamma} \sqrt{6 \left( \frac{4\alpha}
{\gamma}-1 \right)}\,\, {\rm ln}\, t\,, \\
& &  \xi(\phi)=\frac{\alpha}{2c_1}
\exp \left( \frac{2\gamma \phi}{\sqrt{6(4\alpha/\gamma-1)}}
\right)\,.
\ea

Let us next derive the spectral indices of scalar and tensor
perturbations. {}From Eq.~(\ref{cRdef}) we find
\ba
\label{exp1}
c_{\cal R}^2=\frac{(2-\gamma)(7-6\gamma+\gamma^2)}
{5-4\gamma+\gamma^2}\,,
\ea
which is positive for $0<\gamma<1$.
Since $\epsilon_1=-\epsilon_2=\gamma, \epsilon_3=0$ and
$\epsilon_4=0$ in Eq.~(\ref{epdef}), we obtain
\ba
n_{\cal R}-1=3-\left| \frac{3-\gamma}{1-\gamma} \right|\,.
\ea
When $0<\gamma \ll 1$ this reduces to
$n_{\cal R} \simeq 1-2\gamma$,
thereby giving a nearly scale-invariant curvature perturbation.

Meanwhile the variable $c_{T}^2$ for the tensor
perturbation is given by
\ba
\label{exp2}
c_T^2=2\gamma -5\,,
\ea
which is negative for $0<\gamma<1$.
This leads to strong negative instabilities for small-scale
tensor perturbations invalidating the assumption of
linear perturbations.
Note that this type of instabilities has been also
found in Ref.~\cite{Calcagni} (see also Ref.~\cite{DeFelice}).
Unless the system is initially in a Minkowski vacuum state
(in which $c_T^2$ is positive) before the GB term becomes
important, a problem arises when we quantize tensor modes.
It is interesting to note that the graviton is subject
to this severe ultraviolet instability when we impose the condition
to realize inflation.

Instead of the power-law solution (\ref{hp}), one may
search for de Sitter solutions where $H$ is a constant.
{}From Eqs.~(\ref{s1}) and (\ref{s2}), we get
\ba
& & \dot{\xi}=\frac{3}{10c_1 H}+Ae^{-5H t}\,,\\
& & \omega \dot{\phi}^2=-6H^2 \left(1/5
+4c_1 H Ae^{-5H t} \right)\,,
\ea
where $A$ is an integration constant.
Considering the asymptotic regime $e^{-5Ht} \to 0$,
$\omega$ is required to be negative. We also find
\ba
c_{\cal R}^2=14/5\,,\quad c_{T}^2=-5\,.
\ea
These are obtained by taking the limit $\gamma \to 0$
in Eqs.~(\ref{exp1}) and (\ref{exp2}).
Thus the de Sitter solution can be regarded as the special
case of the power-law solution corresponding to the
limit $\gamma \to 0$.

\subsection{Models with $F(\phi) \sim e^{\mu \phi}$}

We now consider the case in which one of
the exponential terms dominates in the function $F(\phi)$, i.e.,
$F(\phi)=e^{\mu \phi}$.
For the tree-level action with $F(\phi)=-\omega(\phi)=e^{-\phi}$
and $\xi(\phi)=\lambda e^{-\phi}$, it is known that
there exists a de Sitter solution characterized by
constant $H$ and $\dot{\phi}$ \cite{higherpapers,CHC01}.
In what follows, we search for de Sitter solutions for the
more general coupling $F(\phi)=e^{\mu \phi}$ and evaluate
the spectral indices of scalar and tensor perturbations.

Since $H^2$ is a constant in Eq.~(\ref{f1}),  it is natural to find
solutions in which each term on the r.h.s. of
Eq.~(\ref{f1}) is also a constant.
This then requires that $\dot{\phi}$ should be a constant and that
$\omega (\phi), \xi(\phi)$ have the same dependence as
$F(\phi)$, i.e., $\omega (\phi)=\omega_0 e^{\mu \phi}$ and
$\xi (\phi)=\lambda e^{\mu \phi}$.
Hence from Eqs.~(\ref{f1}) and (\ref{f2}) we obtain
the following equations:
\ba
\label{rela1}
& &-\omega_0 \dot{\phi}^2+\mu H \dot{\phi}
-\mu^2 \dot{\phi}^2 \nonumber \\
& & +2\lambda \left[ 2c_1 \mu H^2 \dot{\phi}
(\mu \dot{\phi}-H) +c_2 \dot{\phi}^4
\right]=0\,, \\
\label{rela2}
& & 6H^2-\omega_{0} \dot{\phi}^2+6H \mu \dot{\phi}
\nonumber \\
& &-3\lambda (8c_1 \mu H^3 \dot{\phi} -c_2 \dot{\phi}^4)=0\,.
\ea
These have a trivial solution $(\dot{\phi}, H)=(0, 0)$.
However there exist a number of de Sitter fixed points
with constant $\dot{\phi}$ and $H$ depending on the
model parameters $\omega_0, \lambda, c_1, c_2$ and $\mu$.
In this case, we have $\epsilon_1=\epsilon_2=0$,
$\epsilon_3=\epsilon_4/2=\epsilon_6=\mu \dot{\phi}/2H$ in
Eq.~(\ref{epdef}). Moreover since $c_{\cal R}^2$ and $c_T^2$
are constants, one can use the formula given in
Eqs.~(\ref{nRex}), (\ref{nTex}) and (\ref{ratiodef}), namely
\ba
\label{nRdi}
&& n_{\cal R}-1=n_T=3-\left| 3+\frac{\mu \dot{\phi}}{H} \right|\,, \\
\label{nrdi}
&& r=
8 \frac{\omega \dot{\phi}^2+\frac{3(\dot{F}+Q_1)^2}
{2F+Q_2}+Q_3}{\left(H+\frac{\dot{F}+Q_1}{2F+Q_2}
\right)^2 \left(F+\frac{Q_2}{2}\right)}
\left( \frac{c_{\cal R}^2}{c_T^2} \right)^{\nu_{\cal R}}\,,
\ea
where we have used $\nu_{\cal R}=\nu_T=|\epsilon_3+3/2|$.
This result is valid if both $c_{\cal R}^2$ and
$c_T^2$ are positive.
If either $c_{\cal R}^2$ or $c_T^2$ is negative,
we confront with ultraviolet instabilities for small-scale perturbations.
Note also that the tensor-to-scalar ratio becomes complex.
Equation (\ref{nRdi}) shows that when
$|\mu \dot{\phi}/H| \ll 1$ one can obtain nearly
scale-invariant spectra of curvature perturbations.

\begin{table*}[t]
\begin{center}
\begin{tabular}{|c|c|c|c|c|c|c|c|} \hline\hline
Name & Parameters &  $(\dot{\phi}, H)$ & $n_{\cal R}$ & $n_{T}$ & $c_{\cal R}^2$ &
 $c_{T}^2$ & $r$ \\
\hline
\hline
(a)  & $\lambda=-1/4, c_1=1, c_2=-1, \mu=1$
& $(-1.40, 0.62)$ & 3.28 & 2.28 & $-7.56$
&  $22.2$ & -- \\
\hline
(b) & $\lambda=-1/4, c_1=1, c_2=-1, \mu=10^{-2}$
& $(-5.94, 22.3)$ & 1.0027 & $2.7 \times 10^{-3}$ & $1.72$
&  $-3.10$ & -- \\
\hline
(c) & $\lambda=-1/4, c_1=1, c_2=0, \mu \ge 1$
& -- & -- & -- & -- &  -- & -- \\
\hline
(d1) & $\lambda=-1/4, c_1=1, c_2=0, \mu=10^{-2}$
& $(-4.56 \times 10^4, 91.3)$ & 2.00 & 1.00 & 1.40
&  $-5.00$ & -- \\
(d2) & $\lambda=-1/4, c_1=1, c_2=0, \mu=10^{-2}$
& $(-11.5, 10.4)$ & 1.011 & 0.011 & 2.89
&  $-5.29$ & -- \\
\hline
(e1) & $\lambda=1/4, c_1=1, c_2=-1, \mu=1$
& $(1.22, 1.48)$ & 0.174 & $-0.826$ & 2.32
&  0.611 & 46.4 \\
(e2) & $\lambda=1/4, c_1=1, c_2=-1, \mu=1$
& $(1.46, 1.35)$ & $-0.079$ & $-1.079$ & $-0.819$
&  1.162 & -- \\
\hline
(f1) & $\lambda=1/4, c_1=1, c_2=-1, \mu=10^{-2}$
& $(-1.41, 0.41)$ & 1.034 & 0.034 & $-1.2 \times 10^{-3}$
&  0.994 & -- \\
(f2) & $\lambda=1/4, c_1=1, c_2=-1, \mu=10^{-2}$
& $(1.4158, 0.4042)$ & $0.9650$ & $-0.0350$ & $1.20 \times 10^{-3}$
&  1.006 & $7.1 \times 10^{-3}$ \\
\hline
(g1) & $\lambda=1/4, c_1=1, c_2=-1, \mu=10$
& $(-1.95, 1.00)$ & $-12.4$ & $-13.4$ & $0.19$
&  $-18.4$ & -- \\
(g2) & $\lambda=1/4, c_1=1, c_2=-1, \mu=10$
& $(1.84, 1.00)$ & $-17.4$ & $-18.4$ & $0.17$
&  19.3 & -- \\
(g3) & $\lambda=1/4, c_1=1, c_2=-1, \mu=10$
& $(0.14, 1.42)$ & $9.7 \times 10^{-3}$ & $-0.99$ & $1.007$
&  0.98 & 5.73 \\
\hline
(h) & $\lambda=1/4, c_1=1, c_2=0, \mu=1$
& $(1.03, 1.64)$ & 0.37 & $-0.63$ & 1.26
&  0.095 & $1.15 \times 10^3$ \\
\hline
(i) & $\lambda=1/4, c_1=1, c_2=0, \mu=10^{-2}$
& $(11.5, 10.5)$ & 0.99 & $-0.01$ & 2.72
&  $-4.74$ & -- \\
\hline
(j) & $\lambda=1/4, c_1=1, c_2=0, \mu=10$
& $(0.14, 1.42)$ & $9.8 \times 10^{-3}$ & $-0.99$ & 1.007
&  $0.98$ & 5.73 \\
\hline
\end{tabular}
\end{center}
\caption[crit]{De-Sitter fixed points $(\dot{\phi}, H)$
for a number of different combinations of $\lambda$, $c_1$, $c_2$ and
$\mu$ with $\omega_0=-1$.
In the case (c) there exists no de Sitter fixed point.
The tensor-to-scalar ratio becomes a complex number
if either $c_{\cal R}^2$ or $c_T^2$ is negative.
}
\label{crit}
\end{table*}

At the tree level, one has $\mu=-1$, $\omega_0=-1$, $\lambda=-1/4$
(for bosonic strings), $c_1=1$ and $c_2=-1$. Then
we get the fixed point $(\dot{\phi}, H)=(1.404, 0.616)$, which
agrees with the result in Ref.~\cite{higherpapers}.
In this case the spectral indices of scalar and tensor perturbations
are highly blue-tilted, i.e., $n_{\cal R}-1=n_T=2.28$.
Moreover we have $c_{\cal R}^2=-7.56$ whereas $c_T^2=22.2$,
which means that the scalar perturbation exhibits
violent negative instabilities on small scales.

In Table I we show de Sitter fixed points $(\dot{\phi}, H)$
together with $n_{\cal R}$, $n_T$, $c_{\cal R}^2$,
$c_T^2$ and $r$ for a number of different model parameters.
Since Eqs.~(\ref{rela1}) and (\ref{rela2}) are invariant under the
simultaneous sign changes of $\mu$ and $\dot{\phi}$,
it is sufficient to investigate the case of positive $\mu$.
Generally if we choose the values of $\mu$ larger than order
unity, the spectral indices deviate from scale-invariant ones
and are incompatible with observations.
When $\mu \ll 1$ it is possible to get $n_{\cal R} \simeq 1$
because $|\mu\dot{\phi}|$ can be much smaller than $H$.
However in such a situation the quantity $c_T^2$ typically
becomes negative as we see in the cases (c), (d) and (i)
in Table I.
This is consistent with the results of constant
$F(\phi)$ obtained in the previous subsection.

There exist exceptional situations which lead to
$n_{\cal R}-1 \simeq n_T \simeq 0$ and $r \ll 1$
with positive $c_{\cal R}^2$ and $c_T^2$.
Such an example is given by the fixed point (f2) in Table I.
We find that this is similar to the case in which the GB term
is absent, i.e., $c_1=0$ and $c_2=-1$.
In fact, when $\lambda=1/4$, $c_1=0$, $c_2=-1$ and $\mu=10^{-2}$,
we have $(\dot{\phi}, H)=(1.4162, 0.4035)$, $n_{\cal R}=0.9649$,
$n_T=-0.0351$, $c_{\cal R}^2=1.44 \times 10^{-3}$, $c_T^2=1$
and $r=9.3 \times 10^{-3}$, whose values are similar to those given
in the case (f2).
Hence the model (f2) is not much different from the kinetic
inflation we discussed in the previous section.
Provided that the GB term is subdominant relative to the
$c_2 (\nabla \dot{\phi})^4$ term, it is possible to realize
the observationally supported density perturbation with the
suppressed tensor-to-scalar ratio.
However if the GB term dominates over $c_2 (\nabla \dot{\phi})^4$,
we are faced with the ultraviolet instability of tensor perturbations
while nearly scale-invariant spectra of scalar perturbations
are possible. This is a generic property associated
with inflation induced by the GB term.
Table I corresponds to the parameter $\omega_0=-1$,
but we have also examined the case $\omega_0=1$.
We have thus confirmed that it is difficult to satisfy all the conditions
$n_{\cal R} \sim 1$, $n_T \sim 0$, $r \ll 1$ and
$c_{\cal R}^2, c_T^2>0$
if the correction term ${\cal L}_c$ is dominated by the GB term.

\section{Conclusions}
\label{conclusion}

In this paper we have discussed the possibility to obtain inflationary
solutions and observationally supported density perturbations
for the low-energy string effective action given in (\ref{action}).
The important quantities which are directly linked to observations
are the spectral indices $n_{\cal R}$ and $n_T$ together with the
tensor-to-scalar ratio $r$.
Provided that $c_{\cal R}^2$ and $c_T^2$ are positive constants
in perturbation equations, we obtain the power spectra of scalar
and tensor metric perturbations in Eqs.~(\ref{PS}) and (\ref{PT}),
respectively, together with the ratio $r$ given
in Eq.~(\ref{ratiodef}).
The spectral indices of scalar and tensor perturbations
are given by Eqs.~(\ref{nRex}) and (\ref{nTex}), respectively.
Note that these results are valid not only for slow-roll inflation
($|\epsilon_i| \ll 1$) but also for the non slow-roll
models with constant $\epsilon_i$ that often appear
in dilaton gravity.

The action (\ref{action}) is transformed to the Einstein frame action
(\ref{eaction}) by a conformal transformation (\ref{conformal}).
Since both the curvature perturbation ${\cal R}$ and
the tensor perturbation $h$ are invariant under
the transformation in dilaton gravity, it is sometimes
convenient to study perturbation spectra in the Einstein frame
in order to confront with observations.
For the models without both the dilaton potential $V(\phi)$
and the higher-order
correction ${\cal L}_c$, it is not possible to obtain nearly
scale-invariant spectra of density perturbations because the system
is dominated by the kinetic energy of the field.
Even in the presence of the dilaton potential,
except for a specific case in which $V(\phi)$ is
proportional to $F^2(\phi)$, it is required that
the dilaton is effectively decoupled from gravity
together with the existence of a slowly varying
dilaton potential.

This situation changes if the second-order correction
${\cal L}_c$ given by Eq.~(\ref{correction}) is present.
When $c_1=0$, kinetic inflation is
realized in the Einstein frame if the function $K(\phi)$ becomes
negative. In fact this happens for two-derivative perturbative
corrections to the Kahler potential in heterotic string
theory, see Eq.~(\ref{kcon}).
Along the de Sitter solution (\ref{KL}), the spectral index
$n_T$ and the tensor-to-scalar ratio $r$ vanish whereas
$n_{\cal R}-1$ does not.
If the function $L(\phi)$ is constant in Eq.~(\ref{prho}),
we found that the spectrum of the curvature perturbation is
blue-tilted ($n_{\cal R}>1$) in the case where $K(\phi)$
changes its sign from negative to positive.
For general $L(\phi)$ the spectrum is either red- or blue-tilted,
but $n_{\cal R}$ is very close to scale-invariant one if $K(\phi)$ and
$L(\phi)$ are described by constants plus exponential terms.
Moreover, for power-law solutions in which the scale factor
is given by Eq.~(\ref{plaw}), the spectral index of the curvature
perturbation is red-tilted and can be nearly scale-invariant,
consistent with the WMAP3 data.

When the Gauss-Bonnet term is present in addition to the
$(\nabla \phi)^4$ term, we have found a number of situations
in which inflationary solutions are obtained.
First we studied the case of constant $F(\phi)$ in the absence
of the $(\nabla \phi)^4$ term and showed that
power-law inflation is realized in such a case.
Although scale-invariant spectra are generated for
curvature perturbations, tensor perturbations are faced with
ultraviolet instabilities associated with  negative
coefficient $c_T^2$ when inflation occurs.
We have also studied the case where the function $F(\phi)$
is given by $F(\phi)=e^{\mu \phi}$
and showed the existence of de Sitter solutions characterized
by constant $H$ and $\dot{\phi}$.
If $\mu$ is larger than order one, the spectra of density
perturbations deviate from scale-invariant ones.
When $\mu \ll 1$ it is possible to generate nearly scale-invariant
curvature perturbations, but tensor perturbations again suffer from
negative instabilities on small scales if the correction ${\cal L}_c$
is dominated by the Gauss-Bonnet term.
However as long as the $(\nabla \phi)^4$ term
dominates over the Gauss-Bonnet correction, one can avoid this
problem as it happens in kinetic inflation.

The results in our paper tell us that the condition for inflation,
$|\dot{H}/H^2| \ll 1$, is not enough to generate nearly
scale-invariant spectra of density perturbations.
In addition we generally require that the
parameters $\epsilon_i$ defined in Eq.~(\ref{epdef})
are smaller than order unity and also need to check the signs of
$c_{\cal R}^2$ and $c_T^2$ to avoid ultraviolet instabilities.
We have shown that these conditions are satisfied for some
of the models as in the kinetic inflation discussed
in Sec.~\ref{kinetic}.
Other models with higher-order corrections,
can also satisfy these conditions, provided that the GB term
is not dominant.
It is of interest to extend our analysis to
more general string models as given, e.g., in the
works \cite{Ohta}.

\acknowledgments

The work of N.\,O. and Z.-K. G. was supported in part by the Grant-in-Aid
for Scientific Research Fund of the JSPS Nos. 16540250 and 06042.
S.\,T. is supported by JSPS (Grant No.\,30318802).


\end{document}